\documentclass[prl, twocolumn, floatfix, superscriptaddress]{revtex4-1}
\usepackage{amsmath} % AMS Math Package
\usepackage{amssymb}    % Math symbols such as \mathbb
\usepackage{graphicx} % Allows for eps images

\usepackage{color, verbatim, comment} 

\allowdisplaybreaks

\usepackage{soul}
\begin{document}

\title{Flying in a superfluid: starting flow past an airfoil}

\author{Seth Musser}
\affiliation{Department of Physics, University of Chicago, Chicago IL, 60637, USA }
\altaffiliation[Current address:]{ Department of Physics, 77 Massachusetts Ave, Cambridge, MA 02139}
\email{swmusser@mit.edu}
\author{Davide Proment}
\affiliation{School of Mathematics, University of East Anglia, Norwich Research Park, NR47TJ Norwich, UK}
\author{Miguel Onorato}
\affiliation{Dipartimento di Fisica, Universit\`a degli Studi di Torino and INFN, Via Pietro Giuria 1, 10125 Torino, Italy}
\author{William T.M. Irvine}
\affiliation{James Franck Institute and Enrico Fermi Institute,  Department of Physics, University of Chicago, Chicago IL, 60637, USA }

\begin{abstract}
We investigate superfluid flow around an airfoil accelerated to a finite velocity from rest. Using simulations of the Gross--Pitaevskii equation we find striking similarities to viscous flows: from production of starting vortices to convergence of airfoil circulation onto a quantized version of the Kutta-Joukowski circulation.
We predict the number of quantized vortices nucleated by a given foil via a phenomenological argument.
We further find stall-like behavior governed by airfoil speed, not angle of attack, as in classical flows. 
Finally we analyze the lift and drag acting on the airfoil.

%We investigate the development of superfluid flow around an airfoil accelerated to a finite velocity from rest. Using both simulations of the Gross--Pitaevskii equation and analytical calculations, we find striking similarities to viscous flows: from the production of starting vortices to the convergence of the airfoil circulation onto a quantized version of the classical Kutta-Joukowski circulation. Using a phenomenological argument we predict the number of quantized vortices nucleated by a given foil and find good agreement with simulations. The close analogy to viscous flow breaks down when stalling-like behavior is considered: unlike the classical counterpart, the onset of stalling is governed by airfoil speed, not angle of attack. Finally we analyze the lift and drag acting on the airfoil.
\end{abstract}

\maketitle

\begin{figure}
\begin{center}
\includegraphics[width=0.98\columnwidth]{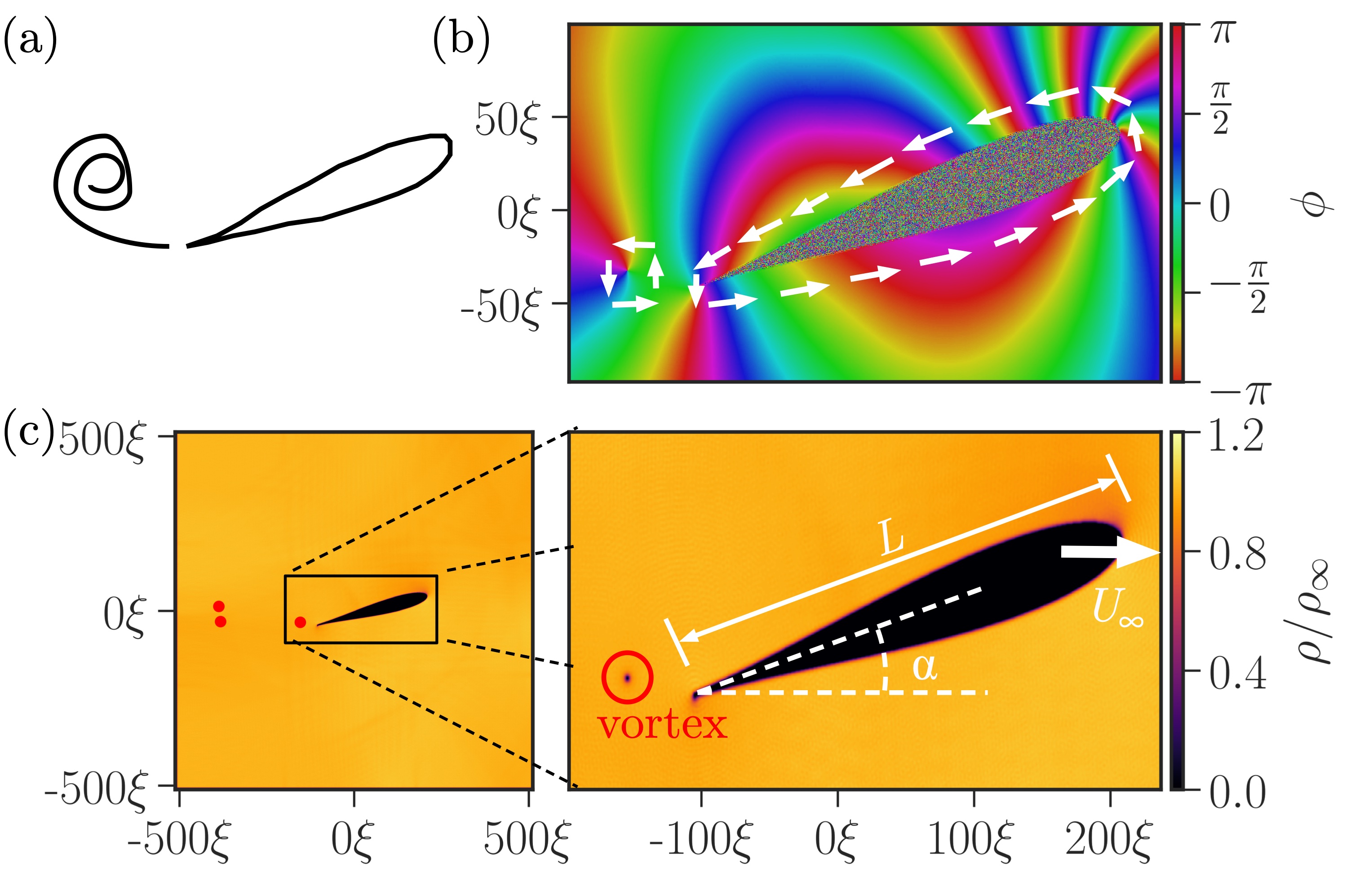}
\caption{
Generation of circulation: 
(a) A cartoon showing the starting vortex produced in a viscous fluid. 
(b) The phase field around the airfoil potential. 
By counting phase jumps around the airfoil the value of the circulation can be obtained. 
A quantized vortex is visible behind the airfoil's trailing edge.
(c) Left hand - the density field in the full computational box.
The density is rescaled by the superfluid bulk density, length scales are expressed in units of $\xi$, quantized vortices are shown as red dots. 
A closer view of airfoil is shown on the right. 
Relevant airfoil parameters are labeled and the vortex is circled in red.}
\label{typ_sim}
\end{center}
\end{figure}

The development of flow around  an airfoil, see sketch in Figure \ref{typ_sim}(a), is a textbook problem in fluid mechanics \cite{d._j._acheson_boundary_1990, anderson_jr._fundamentals_2010, kundu_compressible_2016}. 
Describing this fundamental process has practical relevance since it provides a route to understanding the controlled production and release of vorticity from asymmetric structures. 
In viscous weakly compressible fluids, in the subsonic regime, this release occurs through a subtle interplay of inviscid and viscous dynamics. 

To address the inviscid, incompressible and two-dimensional dynamics, one can use the celebrated conformal Joukowski transformation to relate the flow around an airfoil to the simpler flow past a cylinder.
This makes it possible to readily derive a family of allowed flows, characterized by the value of the circulation $\Gamma$ around the airfoil. 
All but one of these flows feature a singularity in the velocity at the trailing edge. 
To avoid this singularity, the Kutta--Joukowski condition prescribes a circulation, $\Gamma_{KJ} = -\pi U_\infty L \sin(\alpha)$, where $L$ is the airfoil chord, $U_\infty$ the speed and $\alpha$ the angle of attack.
It then follows that the airfoil experiences a lift force per unit of wingspan given by $-\rho U_\infty \Gamma_{KJ}$ and will not experience any drag force.

A major issue with this inviscid theory is that the circulation  $\Gamma_{KJ}$ is prescribed by hand.
Replacing the ideal fluid with an incompressible but viscous fluid and enforcing the no-slip boundary condition gives rise to a boundary layer where the velocity interpolates from zero, on the surface of the airfoil, to the potential velocity  outside \cite{d._j._acheson_boundary_1990}. 
Far from the boundary layer, the flow remains similar to the inviscid case. 
As the trailing edge is approached, the high speeds create a pressure gradient that pulls the boundary layer off the airfoil and into a starting vortex, generating a circulation $\Gamma_{KJ}$ around the airfoil (see Fig.~\ref{typ_sim}(a)). 
Because the airfoil acquires the same circulation as in the ideal case, its lift remains unchanged, though the airfoil experiences a nonzero drag due to viscosity \cite{d._j._acheson_boundary_1990}.

In this Letter we  address the physics of flow past an airfoil in a superfluid. 
In particular, we ask whether (i)
 there exists a mechanism allowing for the generation of a circulation; if so, (ii) whether  the Kutta--Joukowski condition holds and finally, (iii) whether the airfoil experiences  lift and/or drag.
In order to answer these questions we combine an analytical approach with numerical simulations. 
As a model for the superfluid, we consider the Gross--Pitaevskii equation (GPE) which has been successfully used to reproduce aspects of both inviscid and viscous flow, including: the shedding of vortices from a disk \cite{t._frisch_transition_1992, huepe_scaling_1999, jackson_vortex_1997, winiecki_vortex_2000}, an ellipse \cite{stagg_quantum_2014, stagg_classical-like_2015}, a sphere \cite{t._winiecki_motion_2000} and a cylinder \cite{nore_subcritical_1999, stagg_quantum_2014}, the formation of Von K\'{a}rm\'{a}n vortex sheets \cite{huepe_scaling_1999, kazuki_sasaki_benard-von_2010}, the emergence of a superfluid boundary layer \cite{stagg_superfluid_2017}, the dynamics and decay of vortex loops and knots \cite{proment_vortex_2012, kleckner_how_2016}, and the appearance of classical-like turbulent cascades \cite{nore_decaying_1997,kobayashi_kolmogorov_2005}.

%{\it The superfluid model.}

\begin{figure*}
\begin{center}
\includegraphics[width=1.9 \columnwidth]{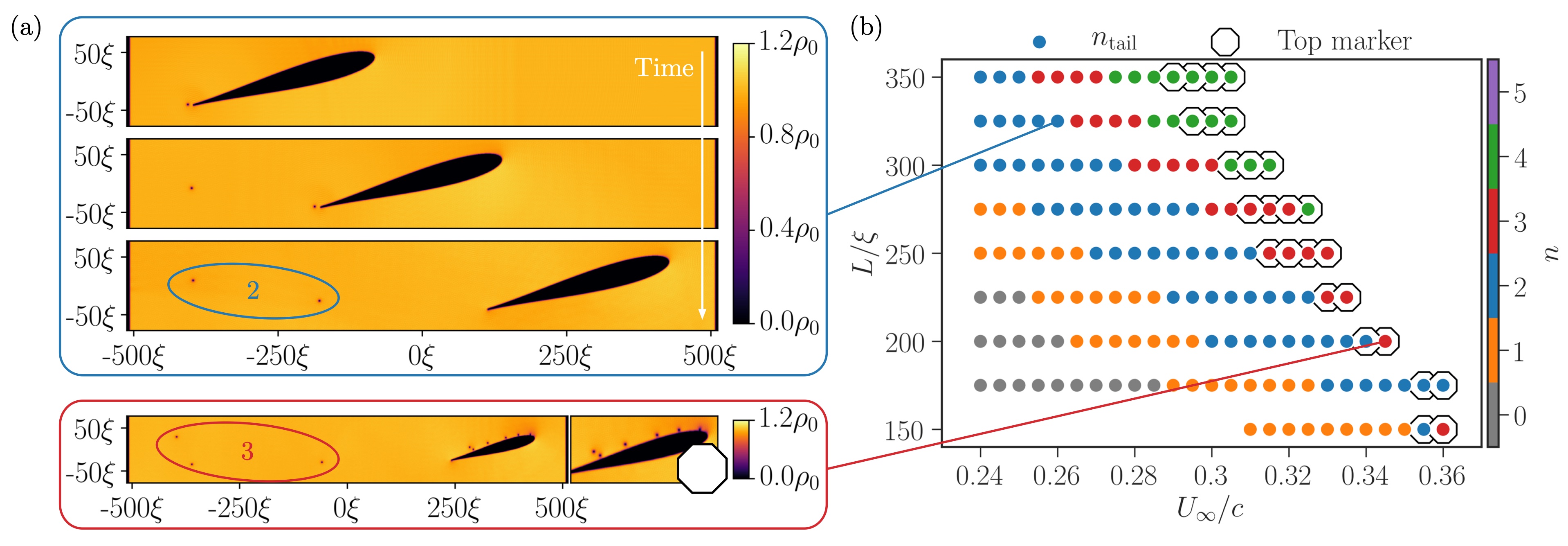}
\caption{Vortex emission: 
(a) Vortices nucleated at the tail and top of a Joukowski airfoil having $\alpha = 15^{\circ}, \tilde{\lambda} = 0.1$. Here the width of the foil scales like $\tilde{\lambda}L$ (see SI). The computational box size is $1024 \xi \times 1024\xi$.
Tail number reflected by color, top nucleation by octagon mark.
(b) Top three frames are snapshots of the density field for $U_\infty = 0.260c$ and $L=325\xi$; here $n_{\mathrm{tail}}=2$. 
Bottom frame has $U_\infty = 0.345c$ and $L=200\xi$. This foil nucleates thrice from tail before nucleating uncontrollably from top; right shows closer view of top vortices.}
\label{results}
\end{center}
\end{figure*}

The two-dimensional GPE is:
\begin{equation}
i\hbar \frac{\partial \psi}{\partial t} = \left(-\frac{\hbar^2}{2m} \nabla^2 + V + g|\psi|^2\right) \psi,
\end{equation}
where $\psi=\psi(x, y, t)$ is the wave-function of the superfluid, $ \hbar $ is the reduced Planck's constant, $ g $ is the effective two-dimensional two-body coupling between the bosons of mass $ m $ and $ V $ is an external potential.
Relevant bulk quantities are the speed of sound $c = \sqrt{g\rho_\infty/m}$ and the healing length $\xi = \sqrt{\hbar^2/(2mg\rho_\infty)}$, with $\rho_\infty$ the superfluid number density at infinity. 
The healing length is the length-scale for the superfluid to recover its bulk density value away from an obstacle; the speed of sound is the speed of density/phase waves of scales larger than $ \xi $.

To understand the superfluid's dynamics in terms of hydrodynamic variables, we make use of the Madelung transformation $\psi = \sqrt{\rho}e^{i\phi}$.
This recasts the GPE into %the following 
hydrodynamical equations for the conservation of mass: $\frac{\partial \rho}{\partial t}+\nabla \cdot (\rho \mathbf{u}  )=0$ and momentum:
\begin{equation}
\frac{\partial \mathbf{u}}{\partial t}+(\mathbf{u} \cdot \nabla)\mathbf{u} = \nabla\left[-\frac{g}{m} \rho +
\frac{1}{m} V+\frac{\hbar^2}{2m^2}
\frac{\nabla^2 \sqrt{\rho}}{\sqrt{\rho}} \right] \, ,
\label{momentum}
\end{equation}
where the density and the velocity of the superfluid are $\rho=|\psi|^2$ and $\mathbf{u} = (\hbar/m)\boldsymbol{\nabla}\phi$, respectively. 
These equations are equivalent to the barotropic Euler equations for an ideal fluid, with the exception of the presence of the quantum pressure term (the last in eq. (\ref{momentum})), negligible at scales larger than $\xi$. 
Circulation around a path $C$ is given by  $\Gamma = \oint_C \mathbf{u}\cdot \mathrm{d}\mathbf{l} = \hbar \Delta \phi/m$, where $\Delta \phi$ is the increment in $\phi$ around $C$: $\Delta \phi$ is quantized in units of $2\pi$ and so is the circulation, in units $ \kappa=h/m $.
Quantized vortices are defined as those points for which the density is zero and the phase winds by $2\pi$ around them.
For example, a vortex  can be seen  in the phase field in Fig.~\ref{typ_sim}(b); the same vortex  also appears circled in red in the density field of Fig.~\ref{typ_sim}(c).

To mimic the motion of an airfoil we add a  potential $V=V[x(t),y]$ moving with  velocity $\dot{x}(t)$ along the $x$ direction.
Within the airfoil shape, the potential has a constant value fifty times higher than the superfluid chemical potential $ \mu=g \rho_\infty $, and decays to zero within a  healing length outside.
At the beginning of each simulation the  potential is  accelerated up to a final velocity, $U_\infty$ which is then kept constant. 
See SI (Supplementary Information) for details of the numerical scheme.

Soon after the airfoil is set into motion, a vortex is nucleated from the trailing edge, much like the starting vortex emitted in classical fluids. 
Our typical airfoil nucleates more than once; the bottom of Fig.~\ref{results}(a) displays an example where three vortices are nucleated from its trailing edge. 
The number of vortices emitted depends in general on the airfoil's terminal velocity $U_\infty$ and length $L$, as shown in Fig.~\ref{results}(b). 
While most of the simulated airfoils reach a steady state post-nucleation, in some cases, highlighted with octagons in Fig.~\ref{results}(b), the airfoil begins nucleating from its top after nucleating from the trailing edge.
Once begun, nucleation from the top  continues for the length of the simulation in a manner reminiscent of the stalling behaviour of a classical airfoil flow. 

These results suggest that, just as for real fluids, an airfoil in a superfluid  builds circulation by vortex emission from its trailing edge. 
%By analogy with viscous boundary layer separation, 
A natural candidate for the mechanism underlying vortex emission is the onset of compressible effects at the tail of the foil \cite{t._frisch_transition_1992, natalia_g._berloff_vortex_2004, winiecki_vortex_2000}.
To estimate this we %neglect the initial acceleration stage, and
consider an airfoil moving with constant terminal velocity $ U_\infty $.
At length scales larger than the healing length, quantum pressure is negligible and the problem simplifies to a classical inviscid compressible fluid one.
The usual condition of compressibility is that relative density variations must be larger than relative speed variations: $ |\nabla \rho|/\rho> |\nabla \cdot \mathbf{u}|/\mathbf{u} $ \cite{kundu_compressible_2016}.
In the steady flow and neglecting the quantum pressure, eq. (\ref{momentum}) is nothing but the classical Bernoulli equation; $ \rho(u)=\rho_\infty+m (U_{\infty}^2-u^2)/2g $, where $\rho_\infty$ is the far field density and $U_\infty$ is the far field velocity in the foil's frame. 
Plugging $\rho(u)$ into the compressibility condition, one obtains that compressibility effects arise when %the local superfluid speed $ u $ is large enough so that
\cite{c._f._barenghi_vortex_2001}:
\begin{equation}
\frac{3}{2}\frac{u^2}{c^2}- \frac{1}{2}\frac{U_{\infty}^2}{c^2}-1 > 0,
\label{eq:comprVel}
\end{equation}
i.e. when the local flow speed is greater than the \textit{local} speed of sound. 
In classical fluids, a dissipative shock is formed where supersonic flow occurs.
On the contrary, reaching the compressibility condition in numerous superfluid models leads to the shedding of vortices
%, which prevents the formation of dispersive shocks
\cite{winiecki_vortex_2000, el_two-dimensional_2009, mossman_dissipative_2018}. 
We use this phenomenological criterion to predict the number of vortices that will nucleate.

\begin{figure}
\begin{center}
\includegraphics[width=0.98\columnwidth]{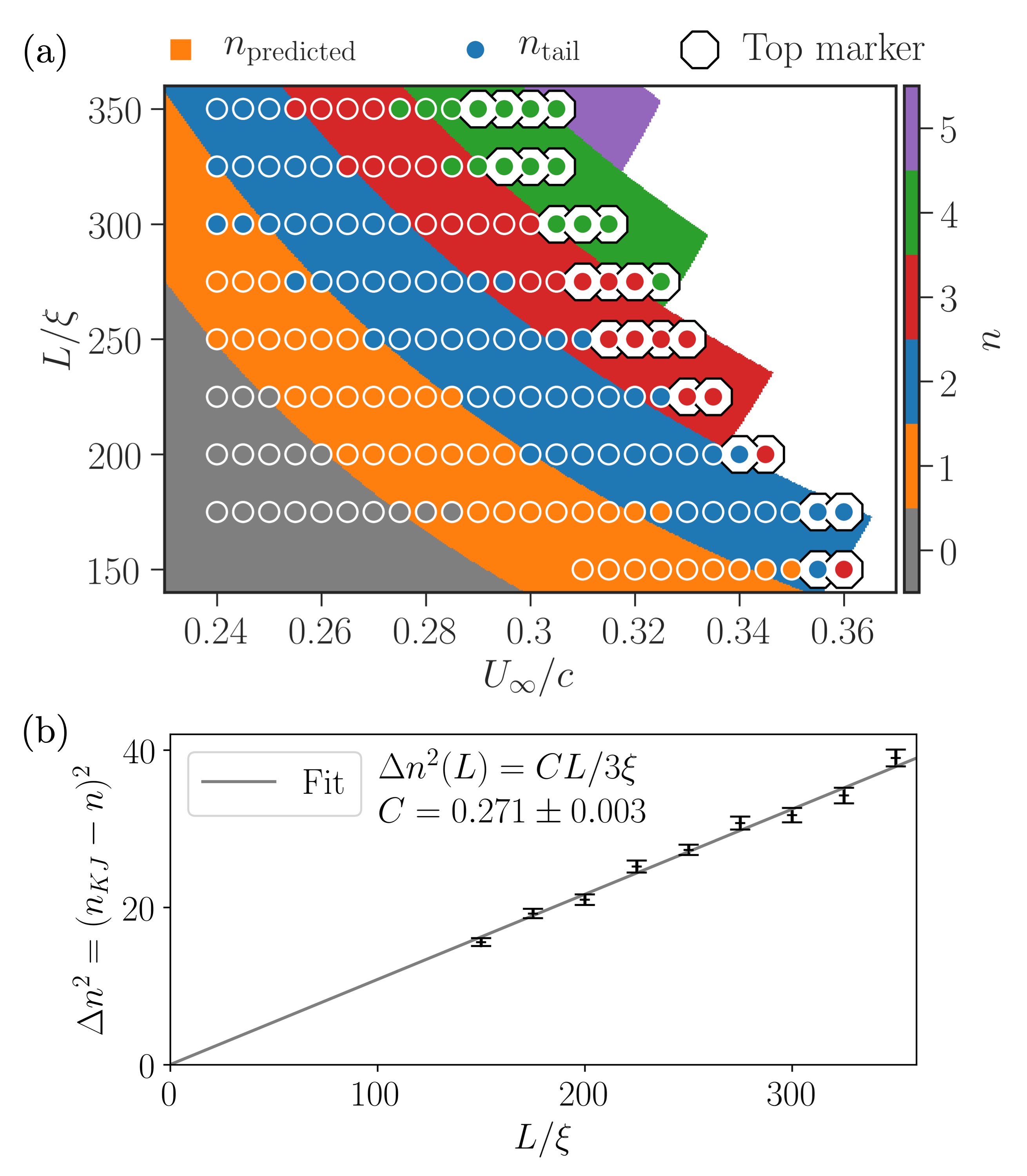}
\caption{Nucleation predictions: 
(a) Plot of tail and top nucleation numbers in $U_\infty-L$ parameter space for $\alpha=15^{\circ}$.
Predictions are stripes in background, white area signifies predicted top nucleation. All predictions used a cut-off distance of $ A = 0.55\xi$ from the foil.
Simulation data is circled in white. 
(b) Values of $\Delta n^2$ calculated for each simulation in (a) with their average plotted vs. $L/\xi$. Errorbars are standard deviation of the mean for the $U_\infty/c$ values in (a).
}
\label{pred_comp}
\end{center}
\end{figure}

We proceed by approximating the velocity of the superfluid $\mathbf{u}$ around the foil by the velocity of an ideal fluid,  $\mathbf{u}_{\mathrm{ideal}}$, around a Joukowski foil of length $L$, terminal velocity $U_\infty$, angle of attack $\alpha$, with a circulation $\Gamma$.
See the SI for a comparison between this approximation and the simulated flow field.
For a circulation $\Gamma \neq \Gamma_{KJ}$, the ideal flow speed $|\mathbf{u}_{\mathrm{ideal}}|$ increases sharply, eventually diverging as the sharp tail is approached.
We expect this divergence to be cut-off by quantum pressure effects arising in the healing layer of size $\xi$. 
Following \cite{rica_s._critical_1993}, we evaluate $\mathbf{u}_{\mathrm{ideal}}$ at a distance $A \xi$, where $A$ is a factor of order unity, and predict vortex nucleation whenever the velocity exceeds the compressibility criterion of eq. (\ref{eq:comprVel}). 
As vortices are nucleated, the value of $\Gamma$ increments accordingly by $\kappa$. 
As $\Gamma$ approaches $\Gamma_{KJ}$ the speeds at the tail decrease and nucleation \textit{from the tail} ends when enough vortices have been emitted to reduce speeds at the tail below the compressibility condition in eq. (\ref{eq:comprVel}). 
We stress that, unlike periodic nucleation of oppositely signed vortices from symmetric obstacles as in \cite{winiecki_vortex_2000, t._winiecki_motion_2000, jackson_vortex_1997, t._frisch_transition_1992, woo_jin_kwon_observation_2016, c._f._barenghi_vortex_2001}, all emitted vortices have the same sign.
Figure~\ref{pred_comp}(a) shows excellent agreement between our simulation data and this prediction for a value $A\sim 0.55 $, close to the value $0.57$ found by Rica et al \cite{rica_s._critical_1993} for a sharp corner. 

\begin{figure}
    \centering
    \includegraphics[width=0.98\columnwidth]{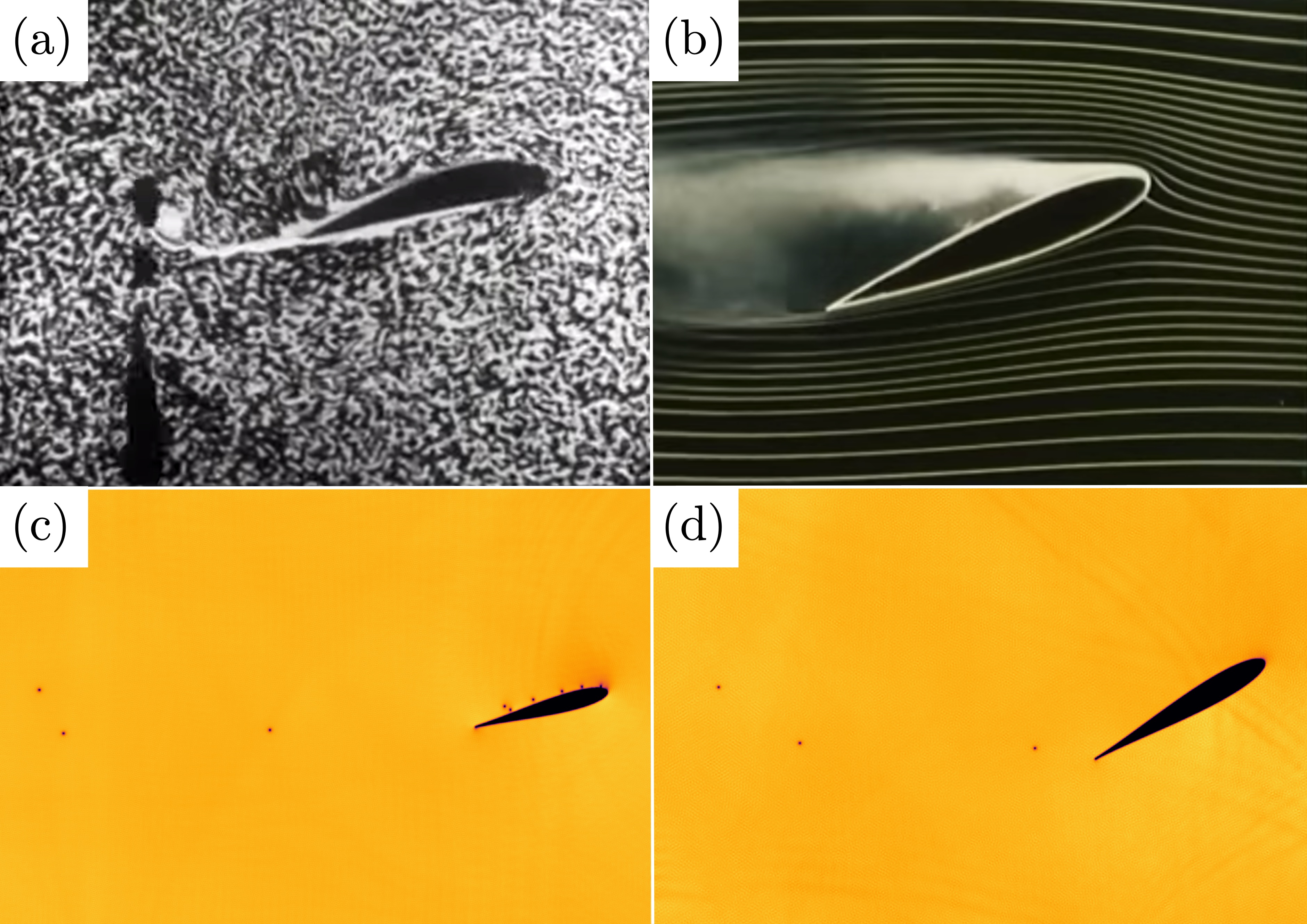}
    \caption{Viscous vs. superfluid flight/stall: (a) Flight of foil in a viscous fluid at a low angle of attack. (b) Stall at high angle of attack. (c) Stall in a superfluid at low angle of attack. (d) Flight at high angle of attack.}
    \label{fig:comps}
\end{figure}

As tail nucleation decreases the speed at the tail, the speed will increase over the top of the foil. 
Once an airfoil has finished nucleating from its tail, if ideal flow speeds at a distance of $A\xi$ from the top are large enough to satisfy (\ref{eq:comprVel}), then we predict the airfoil will stall by continuously emitting vortices from the top. 
The observed stall-like behavior is marked by octagons in Figure \ref{pred_comp}(a); its prediction is represented by the boundary of the colored area.
This marks a radical difference between classical and superfluid flight: stalling in the superfluid is driven by the flow speed at the top of the foil. In viscous flow stalling is primarily a function of $\alpha$. See Figure \ref{fig:comps} \footnote{These images are from the National Committee for Fluid Mechanics Films: Vorticity, Part 2 and Fundamentals of Boundary Layers, copyright 1961 Education Development Center, Inc. Used with permission with all other rights reserved.}.

%\textcolor{green}{DEL: The agreement between the observed stall-like behavior, marked by octagons, and the prediction, represented by the boundary of the colored area of Figure \ref{pred_comp}(a), is not as good as the agreement for tail nucleation. This is connected to the lack of a divergence in ideal flow velocity at the top, which makes the prediction sensitive to detailed differences between inviscid and superfluid flow.}

Returning to tail nucleation, we make our prediction of nucleation number %from the tail 
analytic by appealing to a Taylor expansion of $\mathbf{u}_{\mathrm{ideal}}$ at small distance from the tail. Solving the implicit equation (\ref{eq:comprVel}) for $\Gamma = n\kappa$, reveals that
\begin{equation}
(\Gamma_{KJ}/\kappa - n)^2 \approx C(\alpha)L/(3\xi)
\label{eq:Cpred}
\end{equation}
to first order \footnote{For a better estimation of the healing layer one should in principle extend the expansion to higher order, as suggested for instance in \cite{huepe_scaling_1999} for a cylinder uniformly moving in a superfluid.} (See SI for details). Here $C$ is a constant of order one whose value depends on the angle of attack $\alpha$. If we define $n_{KJ}\equiv\Gamma_{KJ}/\kappa$ to be the number of vortices the foil would nucleate if it acquired a classical circulation, we obtain $\Delta n^2 \equiv (n_{KJ}-n)^2 = C(\alpha)L/3\xi$. We verify this linear relationship by plotting $\Delta n^2$ vs. $L/\xi$ for our simulations, and find excellent agreement shown in Figure \ref{pred_comp}(b).

\begin{figure}
\begin{center}
\includegraphics[width=0.98\columnwidth]{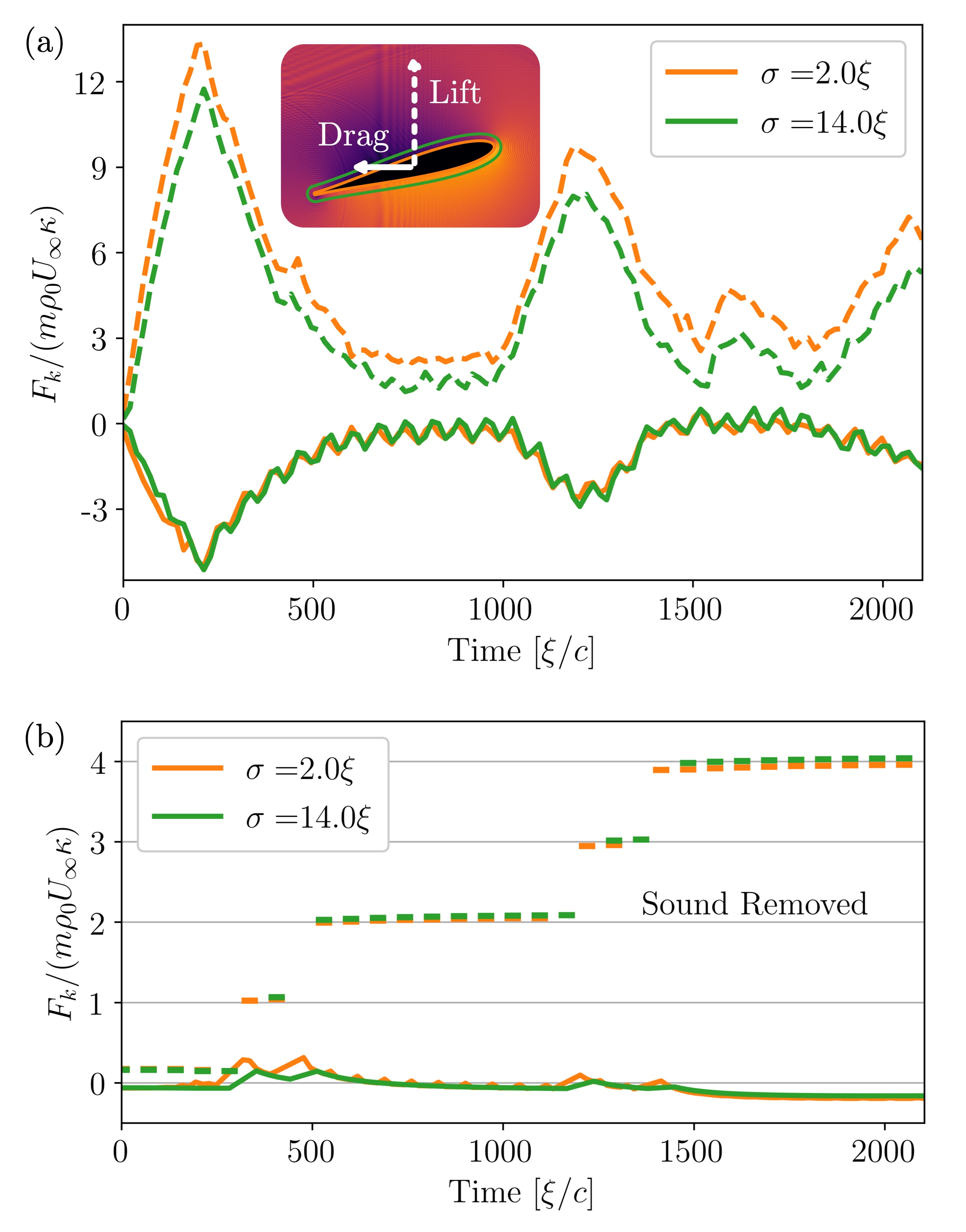}
\caption{Evolution of lift and drag: 
(a) Non-dimensional lift (dotted line) and drag (solid line) experienced by the airfoil throughout simulation with $U_\infty = 0.29c, L = 325\xi, \alpha = 15.0^{\circ},$ and $\tilde{\lambda}=0.1$. 
Inset shows an exaggerated density field around the airfoil. Included are the integration contours for computing the force.
(b) Non-dimensional lift (dotted line) and drag (solid line) experienced by the airfoil using $\mathbf{u}_I$ and $\rho(u_I)$.
A grid is overlaid to demonstrate the quantization of the lift, the steps coincide with vortex nucleation. 
Lift and drag were not computed on a contour if a vortex was within $8\xi$.}
\label{lift_drag}
\end{center}
\end{figure}

Having understood the vortex nucleation, we turn our attention to the force experienced during this process, namely the lift and drag.
The similarity of classical and superfluid vortex nucleation leads us to suspect that an airfoil's lift in a superfluid will be similar to that in a classical fluid, and thus that the Kutta--Joukowski Lift Theorem will nearly hold in a superfluid. 
To calculate the $k$th component of the force exerted by the superfluid on the airfoil one can integrate the stress-energy tensor
\begin{equation}
T_{jk} = m \rho u_j u_k + \frac{1}{2}\delta_{jk}g\rho^2 - \frac{\hbar^2}{4m}\rho \partial_j \partial_k \ln \rho
\end{equation}
around any path $S$ enclosing the airfoil \cite{winiecki_vortex_2000}. 
The results of this calculation for a particular airfoil's simulation are displayed in Fig.~\ref{lift_drag}.
We rescale the computed forces by $m\rho_\infty U_\infty \kappa$, which corresponds to a {\it quantum of lift}: the ideal lift provided by a quantum of circulation. 

The computed lift and drag are clearly not quantized.
We attribute this to transient effects, in particular to the build-up of a dipolar density variation above and below the foil, as can be seen in the inset of Fig.~\ref{lift_drag}(a). 
As discussed in the SI the density dipole, and the  emitted and reflected density wave, lead to contributions to the lift and drag of the same order of magnitude as the two spikes seen in Fig.~\ref{lift_drag}(a).
To %We employ a filter to mathematically 
remove these effects we proceed as follows:%before lift and drag are computed. 
%We expect that 
far from the foil where speeds are low, %and compressibility effects are not important, 
we expect
%unlike at the foil's top and tail, 
that the compressible piece, $\mathbf{u}_C$, of the velocity field will contain only density/sound waves.
As detailed in the SI, the incompressible component of the velocity field  $\mathbf{u}_I \equiv \mathbf{u}-\mathbf{u}_C$, is simply the sum of the ideal velocity field around the foil, $\mathbf{u}_{\mathrm{ideal}}$ and the velocity fields from the emitted vortices.
Replacing $\mathbf{u}$ with $\mathbf{u}_{I}$ and using the density field prescribed by the steady Bernoulli equation, we recalculate the lift and drag and plot it in Fig.~\ref{lift_drag}(b).
Since this calculation differs from that of 
%only difference between this calculation and the calculation of 
lift and drag on an airfoil in ideal fluid %with the same circulation is
only in that  we allowed the density $\rho(u_I)$ to vary in space, it is not surprising that the lift is now quantized and the drag is nearly zero.
%\NOTE{Vinen's classic work \cite{vinen_detection_1961} revealed a similar quantized force in $^{4}$He, though it arose via uncontrolled vortex production, unlike our foil.}

%{\it Conclusions.}
In conclusion we analysed the mechanisms responsible for vortex nucleation from an airfoil and its consequent acquired lift in a two-dimensional superfluid.
On the one hand, we find results reminiscent of the classical theory of airfoils; with the emission of vortices at the trailing edge governed by the elimination of the  singularity predicted by inviscid flow. On the other hand, a marked departure from classical flow is found in the stalling behavior.
Accelerated hydrofoils and wings have recently been used to create vortices of arbitrary shape in classical fluids \cite{kleckner_creation_2013, scheeler_complete_2017},
a technique which might generalize to superfluids, offering a potentially powerful new procedure in superfluid manipulation, vortex generation, and observation of quantized lift -- a measurement originally attempted in $^{4}$He by Craig \& Pellam \cite{PhysRev.108.1109} to demonstrate the quantization of circulation, later detected by Vinen using a different setup \cite{vinen_detection_1961}.
Among the various superfluid experimental realizations, some have recently started to address questions on vortex nucleation and manipulation using moving obstacles including cold atomic gases \cite{woo_jin_kwon_observation_2016, meyer_observation_2017, burchianti_connecting_2018, park_critical_2018, mossman_dissipative_2018, michel_superfluid_2018} and quantum fluids of light \cite{amo_polariton_2011, vocke_role_2016}.
Details of each experimental realization will differ: 3d effects need to be considered for non quasi-two-dimensional BECs, the rotons' emission instead of vortex shedding might be important in $^{4}$He, and out-of-equilibrium  exciton-polariton systems will require  modelling to consider intrinsic forcing and damping terms. The time is right for superfluid flight!

\bibliographystyle{apsrev4-1}

\bibliography{zotero_citation.bib}

\end{document}

\title{Flying in a superfluid: supplementary material}

\author{Seth Musser}
\affiliation{Department of Physics, University of Chicago, Chicago IL, 60637, USA }
\altaffiliation[Current address:]{ Department of Physics, 77 Massachusetts Ave, Cambridge, MA 02139}
\email{swmusser@mit.edu}
\author{Davide Proment}
\affiliation{School of Mathematics, University of East Anglia, Norwich Research Park, NR47TJ Norwich, UK}
\author{Miguel Onorato}
\affiliation{Dipartimento di Fisica, Universit\`a degli Studi di Torino and INFN, Via Pietro Giuria 1, 10125 Torino, Italy}
\author{William T.M. Irvine}
\affiliation{James Franck Institute and Enrico Fermi Institute,  Department of Physics, University of Chicago, Chicago IL, 60637, USA }

\maketitle

\section{Incompressible ideal flow around a Joukowski airfoil \label{sec:foil}}
 
\begin{figure}
\begin{center}
\includegraphics[width=\columnwidth]{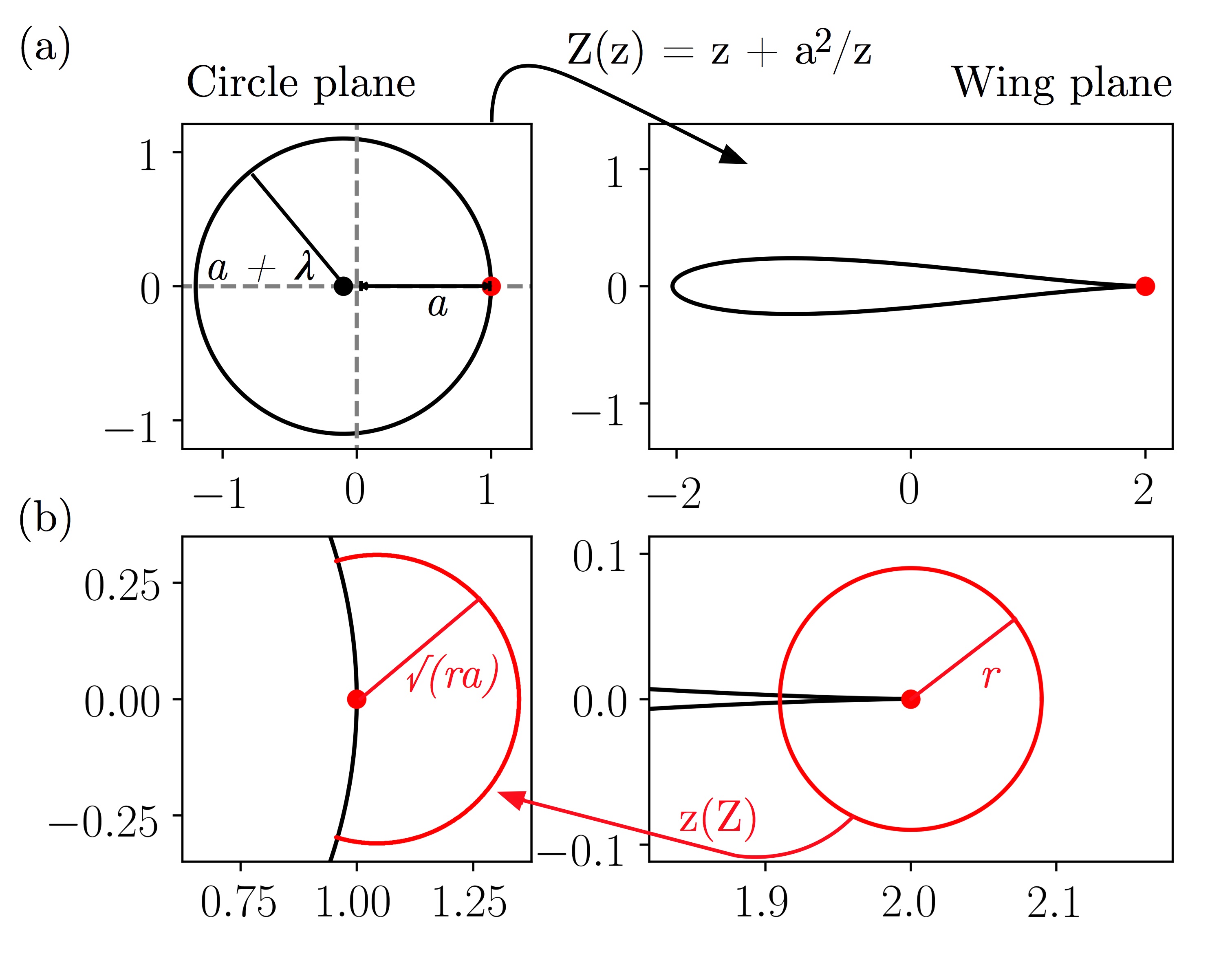}
\caption{Joukowski Map: (a) A demonstration of a circle, with center $z = -\lambda = -0.1$ and radius $a+\lambda = 1.1$, mapping to the symmetric foil under the Joukowski transformation. The point $z=a=1$ in the circle plane maps to the airfoil's tail, $Z=2a$, in the airfoil plane, as shown by the red dot. (b) A demonstration that under the inverse Joukowski map, $z(Z)$, a circle of radius $r$ centered on the tail of the airfoil maps back to a semicircle of radius $\sqrt{ra}$ centered at $z=a$, provided $\sqrt{r/a}\ll 1$. Here $a=1$ and $r=0.09$, so $\sqrt{r/a}=0.3$ was relatively large. Despite this the resulting image is still nearly a semicircle.}
\label{jouk_map}
\end{center}
\end{figure}

In this section we review salient features of the theory of two-dimensional irrotational, incompressible inviscid fluid around an airfoil. In order to generate the airfoil we consider the Joukowski transformation, a conformal map that takes off-center circles to airfoil shapes. It is given by
\begin{equation}
Z(z) = z + \frac{a^2}{z}.
\label{eq:Jmap}
\end{equation}
If we consider the off-centre circle of radius $a+\lambda$ parametrized by $\theta \in [0, 2\pi) $, that is $z(\theta) = -\lambda + (a+\lambda)e^{i\theta}$, its Joukowski transform will be an airfoil whose width will depend on the choice of $ \lambda $.
This procedure is demonstrated in Figure \ref{jouk_map}(a), where we call the $z$-plane the circle plane and the $Z$-plane the airfoil plane, due to this mapping.

The top and tail of the airfoil occur respectively at $ \theta=\pi $ and at $\theta = 0$, or alternatively, at $ z=-(a+2\lambda) $ and at $z=a $.  
Therefore, the Joukowski airfoil has length
\begin{equation}
L = 4a \, \left[1+  \left(\lambda/a\right)^2 \right] + \mathcal{O}\left[\left(\lambda/a\right)^3\right]
\label{length_expansion}
\end{equation}
and width scaling like $3\sqrt{3}\lambda \sim (3\sqrt{3}/4)(\lambda/a)L$ (provided $\lambda  \ll a$) \cite{d._j._acheson_classical_1990}. We denote the non-dimensional $\lambda/a$ as $\tilde{\lambda}$ for use in the main text.

Because Laplace's equation prescribing two-dimensional irrotational, incompressible inviscid flow is invariant under conformal mapping, we can understand this flow around an airfoil by first solving it for a circular impenetrable object and then mapping it to an airfoil via the Joukowski transformation. 
The problem of a circular impenetrable object possessing a circulation $ \Gamma $ and moving in a steady flow with a horizontal velocity at infinity $ U_{\infty} $ can be completely solved \cite{d._j._acheson_classical_1990}.
The solution relies on defining the complex potential $ w = \phi+i\psi $, an analytic function of the complex variable $ z=x+i y $ where $ \phi $ and $ \psi $ are respectively the velocity potential and the stream function of the irrotational incompressible flow.
The velocity components $ u $ and $ v $ are directly computed as
\begin{equation}
\begin{split}
\frac{d w}{dz} = & \,\,  \frac{\partial \phi}{\partial x}+i\frac{\partial \psi}{\partial x}=u-i v \\
 = & \,\, U_\infty \left(1-\frac{a^2}{z^2}\right) - \frac{i \Gamma}{2\pi z} \, .
\label{eq:cyl}
\end{split}
\end{equation}

In the airfoil plane the velocity can be calculated from the complex velocity potential $ W(Z)=w[z(Z)] $ and eq. (\ref{eq:cyl}), where $z(Z)$ is the inverse of the Joukowski transformation. 
The velocity components $ U $ and $ V $, expressed for simplicity in the circle plane variable $ z $, result in
\begin{equation}
\begin{split}
\frac{d W}{dZ} = & \,\, U-i V \\
= & \,\, e^{i\alpha}\frac{U_\infty \left[e^{-i\alpha} - \left(\frac{a+\lambda}{z+\lambda}\right)^2e^{i\alpha}\right] - \frac{i\Gamma}{2\pi (z+\lambda)}}{1-a^2/z^2} \, ,
\label{u_ideal}
\end{split}
\end{equation}
where $ \alpha $ is the angle of attack of the airfoil with respect to the horizontal uniform flow at infinity $ U_\infty $. Then $\mathbf{u}_{\mathrm{ideal}}=U\hat{\textbf{x}}+V\hat{\textbf{y}}$.

\section{Analytic Estimation of Nucleation Number}

\begin{figure}
\begin{center}
\includegraphics[width=\columnwidth]{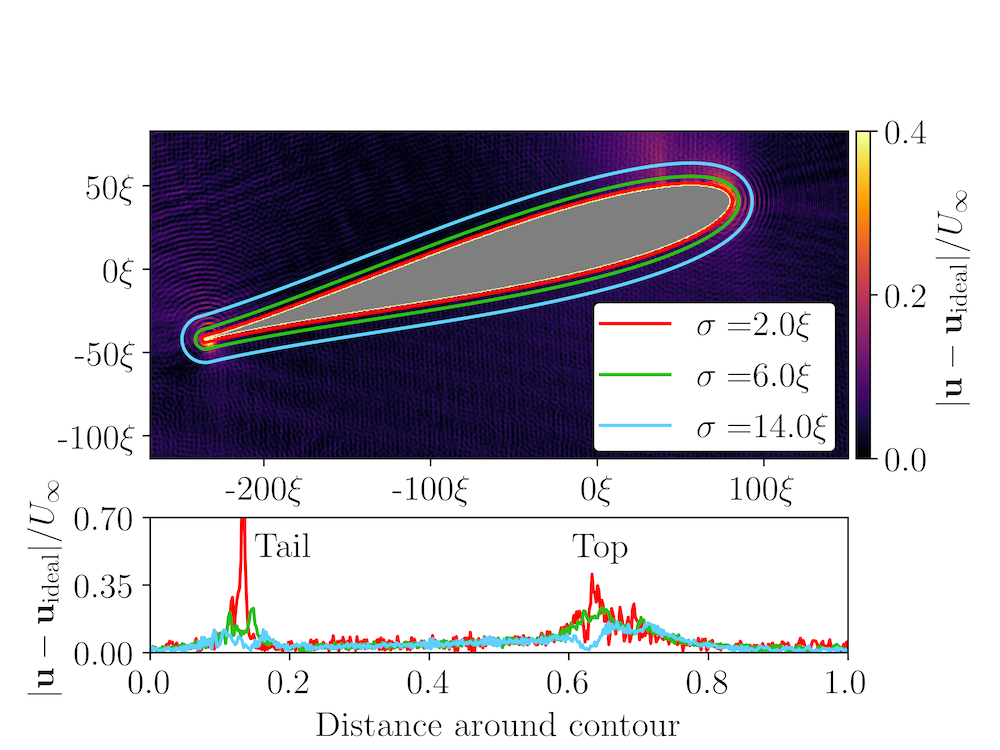}
\caption{Approximation Justification: Demonstration that $\mathbf{u}$ approaches $\mathbf{u}_{\mathrm{ideal}}$ with increasing distance from airfoil. Sampling $|\mathbf{u}-\mathbf{u}_{\mathrm{ideal}}|/U_\infty$ on equidistant contours around the airfoil reveals the difference is largest near the tail and top of the airfoil. }
\label{fig:vel_approx}
\end{center}
\end{figure}

In the main text we began our phenomenological prediction of vortex nucleation from the foil by approximating the velocity of the superfluid $\mathbf{u}$ around the foil by the velocity of an ideal fluid,  $\mathbf{u}_{\mathrm{ideal}}$, around a Joukowski foil of length $L$, terminal velocity $U_\infty$, angle of attack $\alpha$, with a circulation $\Gamma$. Though this is indeed an approximation, Figure \ref{fig:vel_approx} reveals it is a relatively sound one, at least a couple of healing lengths from the foil.

In the main text we predicted that a foil would nucleate when $\mathbf{u}_{\mathrm{ideal}}$ breached the compressibility condition 
\begin{equation}
\frac{3}{2}\frac{|\mathbf{u}_{\mathrm{ideal}}|^2}{c^2}  > \frac{1}{2}\frac{U_\infty^2}{c^2} + 1,
\label{comp_cond}
\end{equation}
at a cut-off distance of $\sigma = 0.55\xi$ from the tail. This prediction was done numerically. Here we derive an analytic condition by expanding $|\mathbf{u}_{\mathrm{ideal}}|^2$ on a cut-off circle of radius $r\sim \xi\ll L$ around the tail of the foil in the airfoil plane. We can parametrize the circle by $\theta$, $Z = 2a+re^{i\theta}$. Transforming this back to the circle plane, as shown in Figure \ref{jouk_map}(b), gives us
\begin{equation}
z(Z=2a+re^{i\theta}) = a\left[1+\sqrt{\frac{r}{a}}e^{i\theta/2}+\frac{1}{2}\frac{r}{a}e^{i\theta}+\cdots\right],
\end{equation}
roughly a semicircle or radius $\sqrt{ra}$ centered on $z=a$. Using the expression (\ref{u_ideal}) for $\mathbf{u}_{\mathrm{ideal}}$ and (\ref{length_expansion}) to put $a$ in terms of $L$, we then have
\begin{equation}
|\mathbf{u}_{\mathrm{ideal}}|^2 = \frac{1}{4}\frac{L}{r}U_\infty^2\sin^2(\alpha)\left(1-\frac{\Gamma}{\Gamma_{KJ}}\right)^2 + \mathcal{O}\left(\sqrt{\frac{L}{r}}\right),
\label{first_term_div}
\end{equation}
where $\Gamma_{KJ} = -4\pi(a+\lambda)U_\infty\sin(\alpha) \approx -\pi U_\infty L\sin(\alpha)$ is the value of the circulation for which there is no velocity divergence at the tip, $r=0$.

We want to solve the implicit equation (\ref{comp_cond}) for $\Gamma = \kappa n$, replacing the inequality with an equality, where we use the expansion (\ref{first_term_div}). 
Here $n$ is the number of vortices that we expect the given foil to nucleate.
Since the right hand side of (\ref{comp_cond}) is of order one this means $\Gamma$ must be chosen so that $|\mathbf{u}_{\mathrm{ideal}}|^2/c^2$ is of order one also. 
The expansion (\ref{first_term_div}) then reveals $(1-\Gamma/\Gamma_{KJ})^2 \sim r/L$ for small $r\ll L$. 
We therefore write $(1-\Gamma/\Gamma_{KJ})^2 = f\cdot r/L$, where $f$ is some order one function. 
Substituting this back into $|\mathbf{u}_{\mathrm{ideal}}|^2$ and keeping the first order terms in $r/L$ then makes (\ref{comp_cond}) into self-consistency equation for $f$. 
Solving this for $f = L/r(1-\kappa n/\Gamma_{KJ})^2$ and manipulating will give us the expression
\begin{equation}
\Delta n^2 \equiv (\Gamma_{KJ}/\kappa-n)^2 = \frac{1}{3}C(\alpha,U_\infty,\lambda)\frac{L}{\xi},
\end{equation}
where $C(\alpha,0,0)=r/\xi$ and is slowly varying for the values of $U_\infty$ we considered. 

\begin{figure}
\begin{center}
\includegraphics[width=\columnwidth]{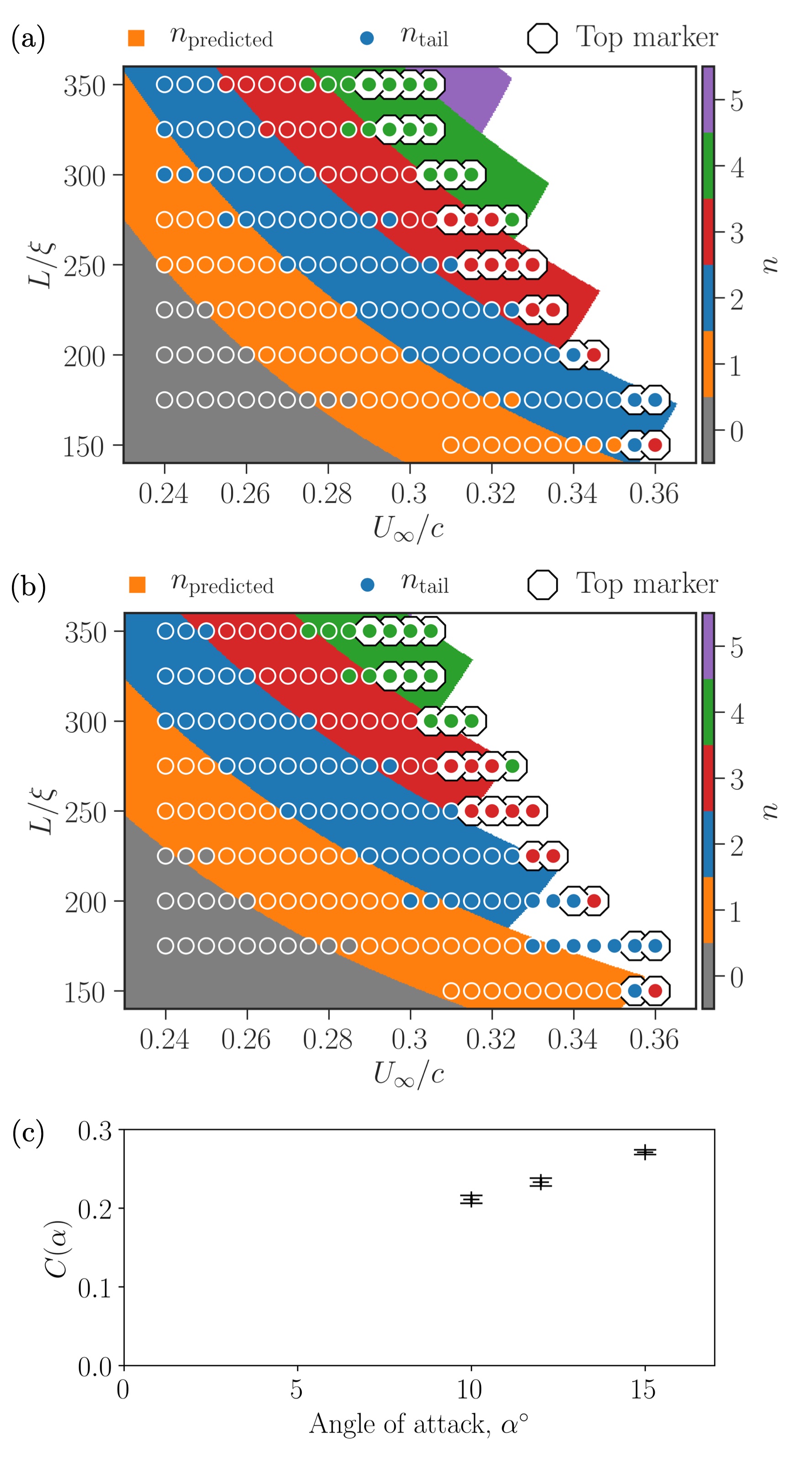}
\caption{Approximation Comparison: (a) Numerically predicted nucleation number vs. simulation nucleation numbers in $U_\infty-L$ parameter space, for $\alpha=15^{\circ}$. A cut-off distance of $A=0.55\xi$ was used to do these numerics. (b) Approximately predicted nucleation numbers also for $\alpha=15^{\circ}$. A fitting value of $C(15^{\circ})=0.271$ was used. (c) A plot of $\alpha$ vs. $C(\alpha)$. A suggestive linear fit is included.}
\label{exact_comparison}
\end{center}
\end{figure}

We treat $C(\alpha,U_\infty,\lambda)\approx C(\alpha)$ as a fitting parameter. For a fixed value of $L$ we simulate a number of foils having different values of $U_\infty$. An example for $\alpha = 15^{\circ}$ can be seen in the ensemble of foil simulations pictured as white outlined dots in Figure \ref{exact_comparison}(a) and (b). For each of these foils we can use the measured number of vortices nucleated from their tail, $n$, to compute $\Delta n^2$ for each $L/\xi$, averaging out $\Delta n^2$'s fluctuations in $U_\infty$. Plotting this averaged $\Delta n^2$ vs. $L/\xi$ then allows us to find the fitting parameter $C(\alpha)$, with its associated error. Such a fit can be seen in Figure 3(b) of the main text.

Figure \ref{exact_comparison} allows us to compare predicted nucleation numbers from this analytic prediction with the more exact numerical scheme discussed in the paper. The stripes in Figure \ref{exact_comparison}(b) were generated by first fitting a value of $C$ from the $\Delta n^2$ vs. $L/\xi$ plot, and then letting $n = \mathrm{round}(\Gamma_{KJ}/\kappa - \sqrt{CL/3\xi})$ be the predicted nucleation number. Note that the rounding restricts $n$ to be an integer, allowing for the discrete jumps from stripe to stripe. The agreement for tail nucleation number is very good between the numerical and analytical methods, with the analytic approximation even reflecting the data slightly better. 

Our predictions for top nucleation in Figure \ref{exact_comparison}(b) were also addressed differently than they were in the numerical scheme of (a). Due to the lack of velocity divergence at the top of the foil making a simple analytic prediction of stalling is difficult. Thus we stuck to numerical methods; if $\Gamma = \kappa n$, for $n$ predicted by the analytic approach, caused speeds at the top to be large enough that they satisfied (\ref{comp_cond}) then we predicted the foil would stall. A cut-off length was not included in this computation, unlike in the numerical approach. Not adding the cut-off length meant we predicted stalling to occur for smaller values of $U_\infty$ and $L$, as can be seen in Figure \ref{exact_comparison}(b), more accurately reflecting the simulation data.

Figure \ref{exact_comparison}(c) shows a plot of the various fitting parameters $C(\alpha)$ for $\alpha = 10^{\circ},12^{\circ},15^{\circ}$. A linear trend might be suggested by the data. However, it is difficult to measure values below $10^{\circ}$, as the foil speeds must be increased in order to see any nucleation at these angles of attach. But increasing the foil speeds means the foils have less time to reach a steady state post-nucleation before they reach the opposing end of the simulation box.

\section{Flight at high angle of attack \label{sec:highaoa}}

\begin{figure}[H]
    \centering
    \includegraphics[width=\columnwidth]{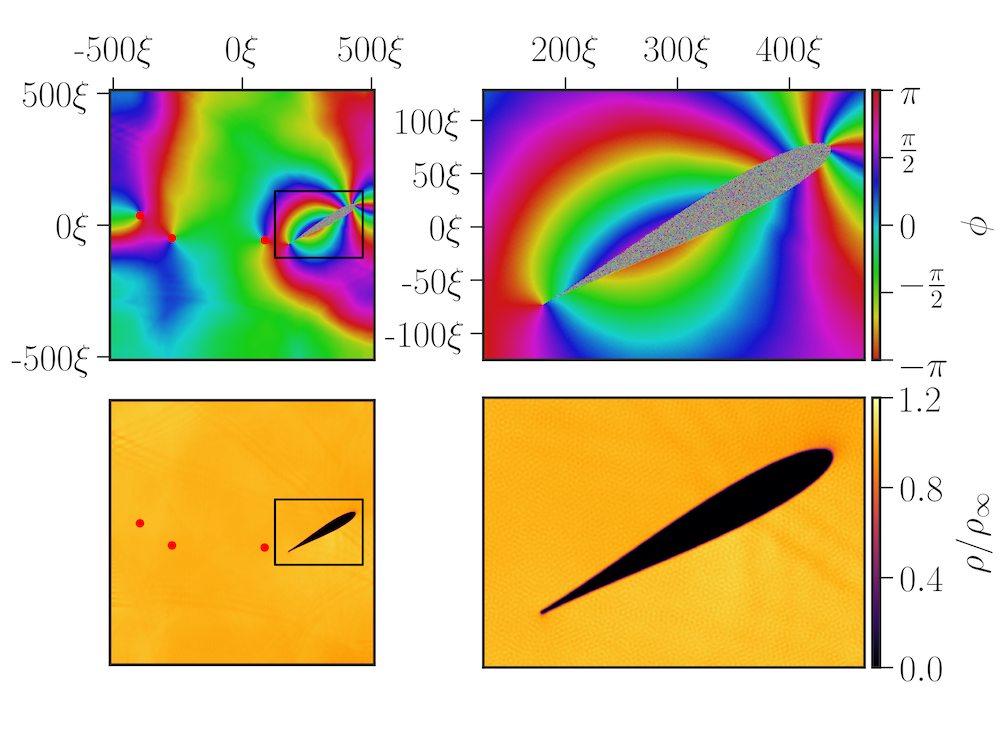}
    \caption{The final phase and density field of a foil with $U_\infty = 0.15c, L = 300\xi, \alpha=30^{\circ}, \tilde{\lambda}=0.1$. The foil has nucleated three times by the time the simulation halts and the vortices are highlighted by red dots in the full phase and density fields on the left-hand side. There is no quantum stalling as no vortices have been nucleated from the top}
    \label{fig:high_aoa}
\end{figure}

In the main text we discussed the possibility of superfluid flight at very high angles of attack, provided the speed of the foil was small enough. An example of this can be seen in Figure \ref{fig:high_aoa}, which displays the phase and density field at the end of a simulation with $\alpha = 30^{\circ}$. The speed of this foil was decreased to $U_\infty = 0.15c$, much smaller than the speeds of the simulated foils at $\alpha = 15^{\circ}$, in order to avoid higher fluid speeds at the top and the consequent quantum stalling that would cause.

We can then see a clear difference between foil flight in superfluids and in viscous fluids as \textit{no} foil is able to fly at this high an angle of attack in viscous fluids without traveling at supersonic speeds. This difference was not as clear when observing nucleation from the tail, as nucleation there is dominated by the inviscid Kutta-Joukowski criterion in both the superfluid and viscous case. In contrast, inviscid flow speeds do not diverge at the top of the foil and thus stalling behavior is instead dominated by boundary layers in the viscous case and by high speeds in the superfluid case. This allows for the strikingly different phenomenon of flight at angles of attack around $30^{\circ}$ in superfluids.

\section{The numerical integration \label{sec:numsim}}

In our simulations we non-dimensionalize the Gross--Pitaevskii equation by rescaling lengths in terms of the healing length $ \xi $, times using $ \xi/c $ where $ c $ is the speed of sound and the superfluid density in terms of the density at infinity $ \rho_\infty $ that we set equal to unity.
We consider a two-dimensional computational box having uniform grid points $2048 \times 2048$ and spacing $ \Delta x = \Delta y = \xi/2 $. 
This spacing is chosen to have the best compromise for the large computational box size $ L_x=L_y=1024 \xi $ and to resolve sufficiently well the healing layer occurring about the quantized vortex core and the about the airfoil external potential.

The airfoil itself has its outline generated in the following way: an off-centred circle is mapped to a symmetric airfoil via the Joukowski transformation (\ref{eq:Jmap}), the resulting airfoil outline is then rotated to an angle of attack $\alpha$ with the horizontal, and finally it is scaled to be the chosen length and placed on the left-hand side of the computational box, with its tail $ 50\xi $ from the left-hand wall and with its centre in the centre of the box's height.
For most of our simulations we took $\alpha = 15^{\circ}$.
The airfoil length is taken between $150\xi$ and $350\xi$, fitting inside the computational box with plenty of room to spare. 
Once the outline is generated we use the python matplotlib package \verb|path| to assign grid points a value of zero if they fall outside the airfoil, and a value of fifty times the chemical potential $\mu = g\rho_\infty$ if they fell inside the airfoil. A small amount of Gaussian smoothing was added to the potential to avoid any sharp edge which may cause fast oscillations (Runge's phenomenon) and cause eventually unwanted sound generation during the evolution: this is done using the Python module $\verb|scikit-fmm|$ to measure distance from the airfoil and then apply Gaussian smoothing with a width of $0.25\xi$. Values larger than this had negligible effects on unwanted sound generation.
We also generate an external confining potential to trap the superfluid into the computational box. 
The box potential has a value of two hundred times the chemical potential at the very edge, and decays to zero $ 5\xi $ away from the edge. 
Its decay is governed by a smooth exponentially-decaying function. 
This potential serves to confine the superfluid, and also reflects any incident sound/density waves.

Having established the external potentials we create the initial wave-function by setting its absolute value to $ \sqrt{\rho} = \sqrt{\rho_\infty} = 1$ outside of the airfoil and edge potential, and $ \sqrt{\rho} = 0 $ inside of the airfoil and edge potential. 
As the GPE is invariant over overall phase translations, the phase field is initially set to zero everywhere, so that the initial wave-function is real valued. 
In order to find the ground-state of the GPE with the airfoil steady, we numerically integrate the GPE forward under imaginary time, keeping the chemical potential fixed.
The numerical scheme for advancing in imaginary time is the same as the one described below to integrate in physical time.
The GPE is first broken into two parts using a standard split-step method \cite{weideman_split-step_1986}: the Laplacian operator $ \mathcal{L} $ is solved exactly in Fourier space, while the nonlinear operator $ \mathcal{N} $ and the external potentials' operator $ \mathcal{V}_a $ and $ \mathcal{V}_b $, corresponding to the airfoil and the confining box respectively, are integrated in physical space.
The time-step $ \Delta t $ is chosen to be smaller than the fastest linear wave period resolved in the computational box, here we chose $\Delta t < 0.1\Delta x^2$.
Assuming periodic boundary conditions and using spectral decomposition the numerical integration can be summarized as
\begin{widetext}
\begin{equation}
\psi(\mathbf{r}, t+\Delta t)= \mbox{FFT}^{-1} \left\{ \mbox{FFT}\left[ \psi(\mathbf{r}, t)  \right]\times \exp(-i \, \tilde{\mathcal{L}}  \, \Delta t ) \right\} \times \exp\left[-i \, (\mathcal{N}+\mathcal{V}_b)  \, \Delta t - i \int_t^{t+\Delta t} \mathcal{V}_a \, dt \right] + \mathcal{O}(\Delta t^2)\, ,
\label{eq:splitstep}
\end{equation}
\end{widetext}
where FFT and FFT$^{-1} $ are respectively the (discrete) direct and inverse Fast Fourier Transforms, $ \tilde{\mathcal{L}} $ is the Laplacian operator represented in Fourier space, i.e. equal to the linear dispersion relation.
Note that due to the splitting, the nonlinear operator $ \mathcal{N} $ is constant within each time-step as is the external box potential operator $ \mathcal{V}_b $; on the contrary the external airfoil potential operator $ \mathcal{V}_a $ is time-dependent if the airfoil moves, hence one has to perform a time integral at each time step (last term in the equation above).  

The numerical integration was performed on GPUs using the Python package $\verb|PyOpenCL|$ to allow access to the $\verb|OpenCL|$ parallel computation API from Python. The Python package $\verb|Reikna|$ was also used for its implementation of the FFT on GPUs. Additionally the function $\verb|locate|$ from the Python package $\verb|trackpy|$ was used to count and track the vortices via the density depletion they caused.

The details of the airfoil operator's time dependence are reported in what follows.
All the airfoils considered in our work were firstly accelerated towards the right-hand side of the computational box as a rigid body with a constant acceleration of $a = c^2/(700\xi)$ until they reached their chosen terminal velocity $U_\infty$. 
This acceleration value was chosen to be large enough so that the airfoils had plenty of room to move at their terminal velocity before nearing the right-hand side of the computational box, but small enough to not cause large sound/density waves or any numerical instability. 
Once the airfoils reached their chosen terminal velocity they moved at this velocity until their top was $75\xi$ from the right-hand side of the computational box, at which point the simulation was halted.
The time integral of the airfoil's external operator in eq. (\ref{eq:splitstep}) becomes remarkably easy noticing that, due to the spectral decomposition, the motion of the airfoil potential results in a simple translation in Fourier space that reads
\begin{equation}
\mathcal{V}_a[x-x(t)] = \mbox{FFT}^{-1}\left\{ \mbox{FFT}(\mathcal{V}_a^0) \times \exp[-ik_xx(t)] \right\} \, 
\label{eq:airfoilFourier}
\end{equation}
where $ \mathcal{V}_a^0 \equiv \mathcal{V}_a(x, y, t=0) $ is the airfoil potential at the initial conditions.
For our purposes:
\begin{equation}
x(t) = \begin{cases} 
\frac{1}{2}at^2 &\mbox{for } 0\leq t\leq U_\infty/a\\ 
U_\infty t - \frac{U_\infty^2}{2a} &\mbox{for } t\geq U_\infty/a
\end{cases}.
\nonumber
\end{equation}
By combining this last expression with eq. (\ref{eq:airfoilFourier}) we can therefore express the time integral in eq. (\ref{eq:splitstep}) in a closed form in terms of complex error functions and exponentials, and integrate in time the GPE with a moving airfoil.

\section{Sound and Helmholtz Decomposition}

In the main text Figure 5(a) shows large spikes in the lift and drag that drown out the contribution assumed to come from the development of circulation via emission of vortices. The supplementary movies demonstrate that the magnitude of the lift and drag is heavily influenced by long wavelength fluctuations in the density field that build-up prior to the initial vortex nucleation and are emitted as a density wave after this nucleation. The first spike in both the lift and drag coincides with such a build-up, while the second spike coincides with the emitted wave's return after bouncing off the walls of the simulation box. We demonstrate that such a density wave will indeed generate spikes in the lift and drag of about the order of magnitude seen in Figure 5(a).

Consider density fluctuations of the form
\begin{equation}
\delta \rho(t,\mathbf{r}) = \epsilon \rho_\infty \exp\left(-\frac{(\hat{\mathbf{n}}\cdot \mathbf{r}' - ct)^2}{2w^2}\right) \ \text{ where } \ \mathbf{r}' = \mathbf{r} - \mathbf{u}_\infty t,
\nonumber
\end{equation}
and $\textbf{u}_\infty$ is the flow field at infinity. These solutions describe a density wave of small amplitude $\epsilon \rho_\infty$ traveling with speed $c$ in the $\hat{\mathbf{n}}$ direction, in the frame of the box. The density wave will have width of about $w \gg \xi$ in the $\hat{\mathbf{n}}$ direction, and will extend the whole length of the box in the $\hat{\mathbf{n}}_\perp$ direction, roughly the form the density wave seen in the movie has when it impinges on the foil after bouncing off the walls of the box. The wavepacket that produces such a density fluctuation will be of long wavelength, i.e. $k\ll 1/\xi$, and is therefore within the linear dispersion range, maintaining its shape during its motion.

The corresponding linearized stress $T_{jk}$ obtained by keeping terms of order $\epsilon$ and dropping the constant zeroth order term gives
\begin{align*}
\delta T_{jk} =& \epsilon m \rho_\infty c^2 A_{jk}\exp\left(-\frac{(\hat{\mathbf{n}}\cdot \mathbf{r}' - ct)^2}{2w^2}\right)\\
A_{jk}\equiv& \left(\frac{\mathbf{u}_\infty}{c}\right)_j\left(\frac{\mathbf{u}_\infty}{c}\right)_k + \left(\frac{\mathbf{u}_\infty}{c}\right)_j\mathbf{n}_k + \mathbf{n}_j\left(\frac{\mathbf{u}_\infty}{c}\right)_k + \delta_{jk},
\end{align*}
where we have dropped terms of $O(\xi^2/w^2)$. The dominant term in $A_{jk}$ is the $\delta_{jk}$ term, which arises due to fluctuations in pressure and is accompanied by Doppler shifting terms that are non-zero but sub-dominant.

To calculate the force we choose the region $\Omega$ to be a large box with side length $L$ that has one of its sides perpendicular to the direction of travel of the pulse, and take the pulse to be incident on this box at the time we compute the force. Then the force becomes
\begin{equation}
F_k \approx \epsilon L m\rho_\infty c^2 A_{1k},
\nonumber
\end{equation}
provided $L^2 \gg w^2$. Since $A_{1k}$ is of order one, this means that we expect the non-dimensional force to be on the order \begin{equation}
\frac{F_k}{m\rho_\infty U_\infty\kappa} \approx \epsilon \frac{L c^2}{U_\infty \kappa} = \epsilon \frac{1}{2\pi\sqrt{2}}\frac{L}{\xi}\frac{c}{U_\infty}.
\label{eq:OrOfMag}
\end{equation}
The magnitude of $\epsilon$ can be eyeballed from the supplementary movie that shows such a density fluctuation, however we estimate its magnitude with a more exact computation aimed at removing sound waves from the velocity and density field.

\begin{figure}
\begin{center}
\includegraphics[width=\columnwidth]{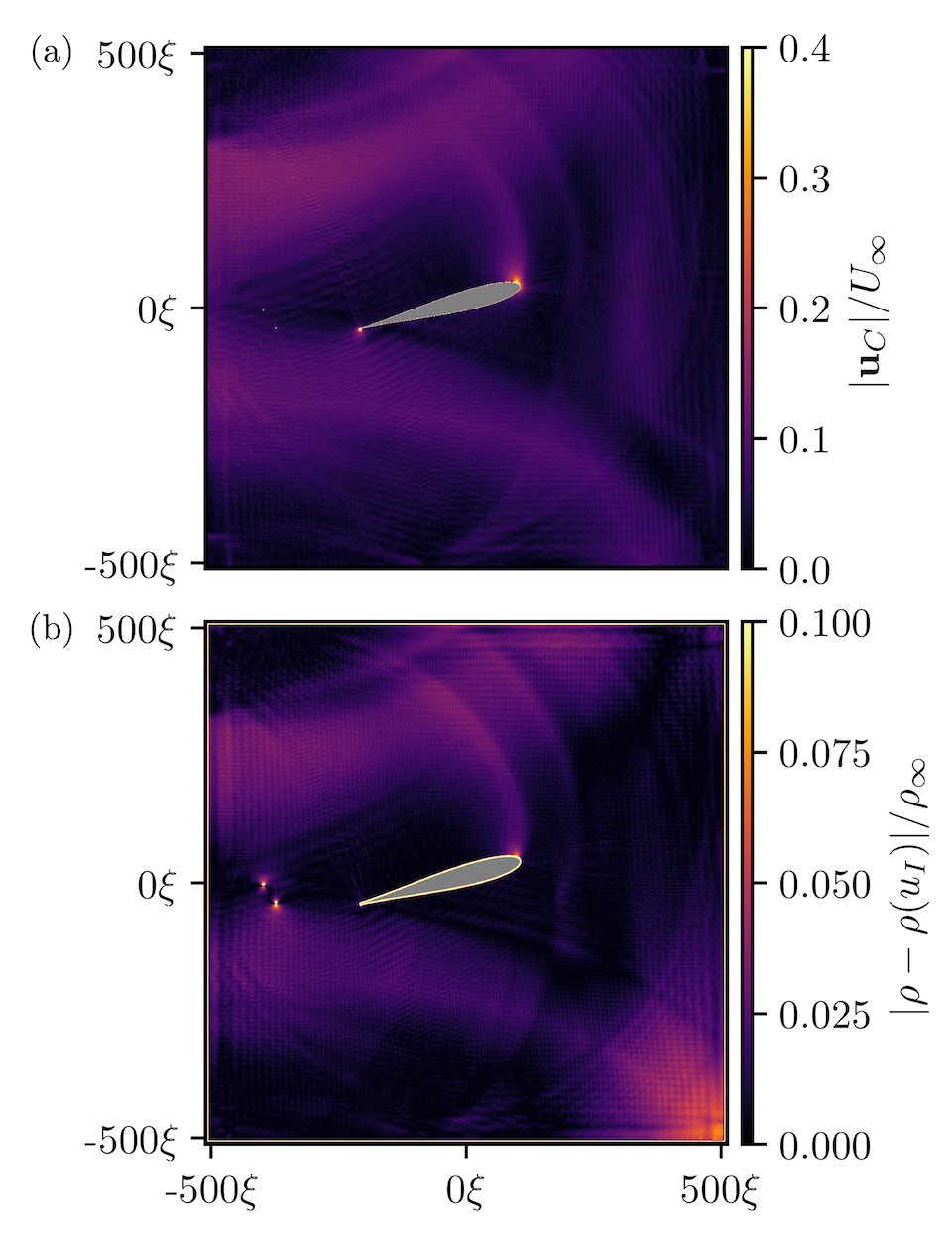}
\caption{Sound waves in velocity and density: (a) The magnitude of the compressible velocity $\mathbf{u}_C$ obtained from subtracting $\mathbf{u}_{\mathrm{ideal}}$ and the point vortex velocity fields from $\mathbf{u}$. The peaks in $|\mathbf{u}_C/|U_\infty$ are due to long wavelength sound waves.  (b) The magnitude of density fluctuations. The density fluctuations are peaked in the same places $|\mathbf{u}_C|/U_\infty$ is, and correspond to the same sound waves.}
\label{sound_fig}
\end{center}
\end{figure}

Though close to the foil's top and tail, where speeds are high, compressibility effects are essential features of the superfluid flow, far from the foil, where speeds are low, the compressibility of the fluid will be due to transient sound wave effects. Thus removing sound waves, at least far from the foil where our contours for computing lift and drag are located, will be mathematically equivalent to removing the compressible parts of the velocity field. We know by the Helmholtz theorem that we can write the superfluid velocity field in the region outside the foil as $\mathbf{u} = \mathbf{u}_I + \mathbf{u}_C$, where $\mathbf{u}_I$ is the incompressible piece and $\mathbf{u}_C$ is the compressible (or curl-free) piece. It is rather simple to find $\mathbf{u}_I$ with reasonable boundary conditions, as we now argue.

In the frame of the foil
\begin{enumerate}
\item[(1)] the velocity field $\mathbf{u}_{\mathrm{ideal}}$ is the unique one that: 
\begin{enumerate}
\item[(a)] is incompressible,
\item[(b)] goes to $U_\infty \hat{\mathbf{x}}$ at infinity,
\item[(c)] gives the foil the right circulation,
\item[(d)] and is such that $\mathbf{u}_{\mathrm{ideal}} \cdot \hat{\mathbf{n}} = 0$ on the surface of the foil, i.e. treats the foil as impenetrable.
\end{enumerate}
\item[(2)] If we now add the vortices and take $\mathbf{u}_I = \mathbf{u}_{\mathrm{ideal}} + \mathbf{u}_{\mathrm{vortices}}$, then $\mathbf{u}_I$ will still satisfy (a),(b),(c) in the frame of the foil. In this case $\mathbf{u}_I \cdot \hat{\mathbf{n}} = \mathbf{u}_{\mathrm{vortices}} \cdot \hat{\mathbf{n}}$, which is not generically zero, but will quickly approach zero as the vortices move away from the foil.
\item[(3)] Writing $\mathbf{u}_I = \mathbf{u}_{\mathrm{ideal}} + \mathbf{u}_{\mathrm{vortices}}$ automatically ensures that $\boldsymbol{\nabla}\times \mathbf{u}_C = 0$, since the only curl in the velocity field outside the foil arose from the point vortices.
\end{enumerate}
Thus we expect that the incompressible piece of the superfluid velocity field outside the foil will be approximated by the sum of the ideal velocity field around the foil and the velocity fields arising from the point vortices. Figure \ref{sound_fig}(a) reveals that the remaining compressible velocity does describe well the long wavelength sound waves that exist in the simulation box.

To remove the soundwaves from the density field as well as the velocity field we consider 
\begin{equation}
\rho(u_I) = \rho_\infty \left(1+\frac{1}{2}\frac{U_\infty^2}{c^2} - \frac{1}{2}\frac{u_I^2}{c^2}\right),
\nonumber    
\end{equation}
which is the density field prescribed by the steady Bernoulli equation assuming a velocity field given by $\mathbf{u}_I$. This allows us to not only calculate the lift and drag with sound waves removed, but also to estimate the magnitude $\epsilon$ of density fluctuations by examining the size of $|\rho - \rho(u_I)|/\rho_\infty$ far from the foil. In the density fluctuation peak visible in the upper left hand corner of Figure \ref{sound_fig}(b) the density fluctuations have a magnitude of about $\epsilon \sim 0.05$. Then from (\ref{eq:OrOfMag}) we expect the non-dimensional force on the foil in Figure 5(a) of the main text to be on the order of $6.5$. This is indeed the same order as the excess second spikes in the lift and drag curves in Figure 5(a). Thus this order of magnitude estimate for the contribution of long wavelength density waves to the lift and drag felt by the foil demonstrates the consistency of our approach for mathematically removing sound waves.

As a final note, both panels of Figure \ref{sound_fig} show two waves passing the front of the foil. These are shock waves that were emitted during both nucleations and they propagate along the foil, later reflecting off the walls of the box. However, their magnitude is much less than the large density wave that builds up during the acceleration of the foil and they do not significantly influence the lift and drag.

\bibliographystyle{apsrev4-1}

\bibliography{zotero_citation.bib}